\documentstyle[12pt,epsfig]{article}
\begin{document}
\title{Higgs Field and Localization Problem.}

\author{ A. Camacho $^{\diamond}$ $^{\circ}$
\thanks{email: abel@abaco.uam.mx}~~and H. Dehnen $^{\diamond}$\thanks{email: dehnen@spock.physik.uni-konstanz.de}\\
$^{\diamond}$University of Konstanz,\\
 P.O. Box 5560, M 678, D-78434, Konstanz, Germany, \\ 
$^{\circ}$Physics Department, \\
Universidad Aut\'onoma Metropolitana-Iztapalapa. \\
P. O. Box 55-534, 09340, M\'exico, D.F., M\'exico.}

\date{}
\maketitle

\begin{abstract} We analyze the role that the excited states of the Higgs field could play in a possible solution to the so 
called localization problem of Quantum Theory. We seek a solution to the aforementioned point without introducing additional fundamental constants or extra hypotheses, 
as has been done in previous works. The electron and Higgs field do indeed have solitonic solutions. 
This last feature renders, in the one-dimensional case, a solution to the localization problem. 
\end{abstract}
\bigskip
\bigskip

\vfil\eject
\section{Introduction.}

\bigskip
\bigskip

Quantum mechanics (QM) is one of the most successful theories in Physics, but very deep conceptual conundrums beset it since its very inception [1]. QM is in perfect agreement with all definite experiments of Physics, but it contradicts our general macroscopic world view [2, 3, 4]. 
The quest of a more general QM theory is already an old one [5], but recently the interest in this search has waxed [6] and has propelled a new momentum to the search of a non-linear QM.

One of the most important puzzles that beset QM is related to the fact that Schr\"odinger equation leads to the unlimited coherent expansion of the center of mass wave function of any isolated system [4]. 
At the root of this problem can be found the clash between the principle of superposition and the experimental evidence which supports the existence of states with far away coherent components only in the case of micro-objects [7].

Let us explain this a little bit better. It is already known that Quantum Physics predicts for a free non-relativistic wave packet, which at certain time is localized in a bounded region, its spreading with the evolution of time. 
If we think of wave packet solutions as possibly the best representation for the free motion of a macroscopic body, then the wave packet corresponding to its center of mass continually widens with time, neverwithstanding, at the same time, experience shows that a macroscopic object has a well defined position. This is the so called localization problem [4].

In this work we show that the Higgs field of elementary particle physics, which is responsible for the mass of the particles, could represent an environment leading to a feasible localization of the center of mass. 
We will not need to introduce new fundamental constants, and the environment, the Higgs field, responsible for the emergence of solitonic solutions, is not introduced by hand, at least at this level. 
In other words, in this context, Higgs field plays a two-fold role, on one hand, elementary particles are endowed by this field with mass, 
and on the other one, these particles show some classical properties as consequence of their interaction with this field.

The new term in Schr\"odinger equation, responsible for the emergence of some interesting properties in the solutions here obtained, might be interpreted as a self gravitational interaction and has the form of Choquard equation, already known from Plasma Physics. 

We will find solitonic solutions, for the electron as well as for the excited states of the Higgs field, for the three and one dimensional cases. Nevertheless, in the latter case the localization property will be restricted to the direction of motion of the electron's wave packet. 
For the former case the localization problem disappears completely and the solitonic velocities of the electron and Higgs field are equal and fixed by the normalization of the electron's wave packet. 
There are some results [8], which seem to point 
out that we might find localized solutions even in the case of three space--like dimensions. 

\bigskip
\bigskip

\section{Motion equations.}

\bigskip
\bigskip
We begin with Dirac equation for an electron and the equation for the excited states of the Higgs field that appear in the Standard Model [9]. After symmetry breaking, we have (with $c $ and $\hbar $ set to unity):

\begin{equation}
 i\gamma^\nu \partial_{\nu}e_l - M[1 + \phi]e_r = 0,
\end{equation}

\begin{equation}
i\gamma^\nu \partial_{\nu}e_r - M[1 + \phi]e_l = 0,
\end{equation}

\begin{equation}
 (\Delta - {\partial^2 \over \partial t^2}) \phi- m^2 \phi  - {M\over v^2}[\bar e_r e_l + \bar e_l e_r] = 0,
\end{equation}

\noindent where $e_l, e_r$ are the left and right parts of the electron's wave function, $M$ and $m$ the electron and Higgs masses, respectively, and $\phi$ the real valued excited Higgs field, 
$v^2$ the expectation value of the vacuum state of the Higgs field; furthermore, equation (3) has been linearized with respect to $\phi$. 
This means, for instance, that those terms appearing in the Higgs potential [9] ($V(\Phi) = -\mu^2\Phi^{\dagger}\Phi + \lambda(\Phi^{\dagger}\Phi)^2$) that are of second or higher order in $\phi$ must be neglected. Upon the excited states of the Higgs field the imposed condition is $\vert\phi\vert^{n+1} < \vert\phi\vert^n$, for any natural number $n$. 
Finally, the interaction with the electro--weak gauge bosons is neglected.
 
Let us now proceed to take the non-relativistic limit of (1). This is done through the introduction of the condition [10] $e_{r,l} = e^{-iMt}E_{r,l}$, $E_{r,l}$ being a function that changes slowly with time, meaning this condition that the energy of the particle is concentrated mainly in its rest mass. 
We may understand this approximation from another point of view, namely we expand the electron field in a series, where the \-pa\-ra\-me\-ter\- used in this expansion is $V_g/c$, $V_g$ being the group velocity of the particle. 
The employed cut off in this series implies that the group velocity of the electron field must be much smaller than the velocity of light.
At this point it is noteworthy to mention that in order to carry out the nonrelativistic limit we must also assume that $|\phi| < M$. This has to be 
done, otherwise the second order term in the corresponding expansion in powers of $(v/c)^2$ could be so large that our nonrelativistic limit would be not valid [11]. 

We may write [10] $E = E_r + E_l$ and $E = \left(\begin{array}{c}\varphi \\ 
\chi\end{array} \right)$. In the nonrelativistic limit, the term $\chi$ is much smaller than $\varphi$, the order of magnitude 
between these two terms goes like $V_g/c$. 

The nonvanishing term $\varphi$ has two components, each one of them associated to one of the two possibles values of the electron's spin. 

Let us denote these components with $\psi$, i. e. $\varphi = \left(\begin{array}{c}\psi_{({1\over 2})} \\ 
\psi_{(-{1\over 2})}\end{array} \right)$. Therefore, the last definition entitles us to write the motion equations, not considering spin, as

\begin{equation}
i {\partial  \psi \over \partial t} + {1\over 2M} \Delta \psi - M \phi \psi = 0, 
\end{equation}

\begin{equation} (\Delta - {\partial^2 \over \partial t^2})\phi - m^2\phi  - {2M\over v^2}\psi^{\ast} \psi = 0.
\end{equation}
From (4) follows a continuity equation for the Schr\"odinger field $\psi$, which therefore can be normalized, if the functions $\psi$ and $\phi$ go to zero sufficiently fast, in the space-like infinity.
 
Assuming in (5) that $\phi$ does not contain time--dependence, namely that it is a static field, we obtain 

\begin{equation}
\phi ({\bf r}) = - {M\over 4\pi v^2}\int {exp[-m\parallel {\bf r} - {\bf r}_2 \parallel ] \over  \parallel{\bf r}  - {\bf r}_2 \parallel} \psi^{\ast}({\bf r}_2) \psi ({\bf r}_2) d^3r_2,
\end{equation}

\noindent and with it equation (4), in the stationary case, goes over onto the integro-differential equation 

\begin{equation}
i {\partial  \psi \over \partial t} + {1\over 2M} \Delta \psi  + {M^2 \over 4\pi v^2} \psi \int{exp[- m \parallel {\bf r} - {\bf r}_2 \parallel ] \over \parallel{\bf r} - {\bf r}_2 \parallel}\psi({\bf r}_2) ^\ast \psi({\bf r}_2) d^3r_2 = 0.
\end{equation}

The integral term can be reinterpreted as a self interaction of gravitational type, and this fact confirms an already known result [12], the excited Higgs field mediates a scalar gravitational-type interaction of short range. 
Commonly, such nonlinear terms are added in an artificial way (see for example: E. Wigner, in Quantum Optics, Experimental, Gravitational and the Measurement Theory, P. Meystre and M. Sally, eds., Plenum Press (1983) or G. Gilburn and C. Holmes, Phys. Rev. Lett. {\bf 56}, 2237 (1986), in which the importance of some cosmic influence is analyzed). In our case they arise automatically from an interaction with a very important field, namely, the Higgs field, which is responsible for the mass of the electron. They do not have to be introduced by hand. We also do not need new extra fundamental constants, as in some previous attempts [13]. 
Equation (7) is the so called Choquard equation, and it has been proved that this equation has stationary normalized solutions in one dimension and in three dimensions for the spherically symmetric case [14], even though, for the three-dimensional case no solution has been found. In this respect, it is noteworthy to mention that concerning the eigenvalue problem nothing is known. 

However, with respect to the localization problem, we search for soliton-like solutions of equations (4)--(5). Through simple substitution it can be checked that the following set of functions is a solution to (4)-(5)

\begin{equation}
 \psi (t, {\bf r}) = {mv \alpha \over \sqrt{2}M^{3\over 2}}sech[\alpha (x-t)]exp[i(\omega t + Mx + \gamma y + \epsilon z)],
\end{equation}

\begin{equation}
\phi (t, {\bf r})= -{\alpha^2 \over M^2}sech^2[\alpha (x-t)],
\end{equation}

\noindent where we have the relation $\alpha^2 = 2M\omega + M^2 + \gamma^2 + \epsilon^2 $ between the constant parameters $\alpha, \omega, M, \gamma$, and $\epsilon$. 
Certainly, it must be mentioned that for the case in which the soliton velocity reaches the speed of light the validity region of the nonrelativistic aproximation may be abandoned. In relation with the localization length, $l$, of this solution it is readily seen that $l\sim 1/\alpha$ and that it is not fixed.

A second solution is given by

\begin{equation}
\psi (t, {\bf r}) = {3m^2v \over 4M^{3\over 2}}{1\over \sqrt{1 - {\mu^2\over M^2}}}sech^2[{m\over 2\sqrt{1 - {\mu^2\over M^2}}} (x-{\mu \over M}t)]e^{[i({2\over M}[{m^2\over 4(1 - {\mu^2\over M^2})} - {\alpha^2\over 4}]t + \mu x + \gamma y + \epsilon z)]},
\end{equation}

\begin{equation}\phi (t, {\bf r}) = -{3\over 4}{m^2\over M^2 - m^2}sech^2[{m\over 2\sqrt{1 - {\mu^2\over M^2}}} (x-{\mu \over M}t)],
\end{equation}

\noindent where we now have the relation $\alpha^2 = \mu^2 + \gamma^2 + \epsilon^2 $. 

Both sets of solutions represent waves traveling along the $x$-axis, which for the particular case $\gamma = \epsilon = 0$ can be normalized with respect to the $x$-axis. 
It is readily seen that they present no time spreading amplitude, i.e. they are solitons, even though, the \-lo\-cali\-za\-tion\- property appears only in the direction of motion. 
In the first case (8)--(9), the soliton velocity is the speed of light, in the second one, (10)--(11), it is $\mu/M$ which implies that $\mu \leq M $. 
 
Noteworthy to mention is the fact that the solution given by equations (10)--(11) is the best one, physically speaking, of the two obtained, because it entitles us to choice any soliton velocity in the interval $[0, 1]$, and in consequence it sets, beforehand no kind of condition to the value for the velocity of the center of mass of a nonrelativistic macroscopic body.

It is not surprising at all that we obtain also solutions that have, in our nonrelativistic limit, no physical meaning. Even in the case of the three dimensional Schr\"odinger equation of a one point-like object of mass $M$, that includes a nonlinear term comprising a non-local self-interacting gravitative term [15], the appearence of nonphysical solutions is unavoidable. 

An important point regarding solutions (10)--(11) concerns the length of the obtained \-lo\-ca\-li\-za\-tion\-. From the aforementioned equations it is clear that the length of the localization is $l \sim {1\over m}\sqrt{1 - {\mu^2\over M^2}}$, 
with other words, it is the Compton length of the Higgs field, $l_H = 1/m$, times the factor $\sqrt{1 - {\mu^2\over M^2}}$, where $\mu$ is a constant and we may choose it such that this localization length becomes $l \leq l_H$, which yields a very small localization region and in consequence a physically reasonable one.

Let us for a moment consider the case of a one-dimensional system. Here we  need to comment that the phrase one-dimensional case means not only that we consider just one space-like dimension in equations (4)--(5), 
but also that the normalization of the wave function is given by the integral $\int_{-\infty}^{\infty}\vert\psi (x)\vert^2 dx = 1$. This last expression has an important consequence, namely $v$ is dimensionless, as can be seen from (5). 

We have, once again, two solutions, in both of them the electron's wave function can be normalized. 

The first set is given by 

\begin{equation}
 \psi (t, x) = {M^{3\over 2}\over \sqrt{2}mv}sech[{M^3\over (mv)^2} (x-t)]exp[iM({M^4 - (mv)^4\over 2(mv)^4} t + x)],
\end{equation}

\begin{equation}
\phi (t, x)= -{M^4 \over (mv)^4}sech[{M^3\over (mv)^2} (x-t)].
\end{equation}

It can be seen that in this case the solitonc velocity is 1, we are once again outside of the nonrelativistic region, and the phase velocity becomes $V_p = (-M^4 + (mv)^4)/(2(mv)^4)$.

The second solution has the form

\begin{equation}
\psi (t, x) = {M^{3\over 2}\over 2mv}sech^2[{M^3\over 3(mv)^2}(x - V_st)]
e^{[i([{2M^5 \over 9(mv)^4} - {M \over 2}V^2_s] t + M V_sx)]},
\end{equation}

\begin{equation}
\phi (t, x) = -{1 \over 3}({M\over mv})^4sech^2[{M^3\over 3(mv)^2}(x - V_st)].
\end{equation}

The solitonic velocity is, $V_s = \sqrt{1- {9\over 4}({m^3v^2\over M^3})^2}$, and in order to remain in the non-relativistic region we must have ${9\over 4}({m^3v^2\over M^3})^2 \simeq 1$. 
The phase velocity is in the case of solutions (14)--(15) $V_p = -{2\over 9}({M\over mv})^4{1\over V_s} + {V_s \over2}$. It is clear that $V_s \geq V_p$.

The expression just derived for the solitonic velocity sets the following bound ${2\over 3}M^3 \geq m^3v^2 \Rightarrow 0.447Mev \geq mv^{2\over 3}$. 
In order to compare with the current experimental bounds for the Higgs particle, we must have an estimation for the expectation value $v$ of the vacuum state of the Higgs field. 
But in our case, a one--dimensional model, the expectation value $v$ can not be the value that this physical \-pa\-ra\-me\-ter\- takes in the three--dimensional 
situation. This last assertion is readily seen if we take a look at the last expression, namely ${2\over 3}M^3 \geq m^3v^2$, $v$ is in the one--dimensional model 
a dimensionless parameter, whereas in the usual Weinberg--Salam theory this \-pa\-ra\-me\-ter\- does have dimensions [9]. 
Therefore, in order to be able to compare the current experimental bounds of the Higgs mass with some type of 
restriction stemming from a model, similar to the one here presented, we must have solitonic solutions of the three--dimensional case. 
 
Clearly, the localization length is $l\sim l_e{m^2 \over M^2}v^2$, where $l_e$ represents the Compton length of the electron. Obviously, ${m^2 \over M^2}>>1$, but $v$ is in the one-dimensional case not fixed and therefore it may be chosen such that ${m^2 \over M^2}v^2 \leq 1$. 

As a matter of fact, if we demand $V_s \in \Re$, and we must set this condition otherwise we can not identify $V_s$ with a velocity, then the following condition must be fulfilled $1 \geq {3\over 2}{m^3v^2\over M^3} \Rightarrow {m^2v^2\over M^2} <<1 \Rightarrow l\leq l_e$, and therefore we obtain through this model a reasonable localization lenght for the electron.  With other words, every physically meaningful solitonic velocity implies a very small localization region.
\bigskip
\bigskip
\section{Conclusions.}
\bigskip
\bigskip
We began with the equations for the electron and excited states of the Higgs-field, this last one linearized, that stem from the Standard Theory. We have also shown that the electron's Schr\"odinger equation contains a new term that might be reinterpreted as a self interaction of gravitational type. 
The form that this equation acquires is already known from Plasma Physics, the so called Choquard equation. In this work the Higgs field can be held responsible not only for 
the emergence of mass terms in some elementary particles, but also for the appearence of some classical properties, i. e.,  the wave function shows no spreading with time.

We found, in the three dimensional case, solutions for the electron and excited states of the Higgs field that in the direction of motion show no spreading with time evolution, but unfortunately, the localization property is restricted to the aforementiond direction. The involved localization length is of the order of magnitude of the Compton length of the Higgs field, a very small one.

For the one dimensional case the localization problem disappears. In this case,  every electronic physically meaningful solitonic velocity implies that its localization region has to have an order of magnitude similar to the electron's Compton length, once again the obtained length is small enough and endowes the model with physical meaning. 
It is readily seen that our solution has the character of a particular solution of the corresponding motion equations, and therefore it can not describe all kind of electron waves. 

The issue of stability and the possible destruction of the soliton solutions by a dissipative environment (structural stability) must be also addressed.

In the case of the Sine-Gordon equation [16], that in the case of one-space-like dimension has the form ${\partial^2 \phi\over \partial x^2} - {\partial^2 \phi\over \partial t^2} = sin(\phi)$ and shows a certain ``similitude'' with the one-dimensional case of (5), it is known that the perturbation of a collision with another soliton is by no means guaranteed to be small. 

The possibility of annihilation of two colliding pulses, for this Sine-Gordon equation, and the existence of different types of inestabilities [17], i. e. convective and nonconvective which depend on the algebraic properties of the involved dispersion relation $\omega(k)$, show that even for a simpler equation than ours the problem of stability is a very complicated one and that the role played by the environment has to be carefully investigated, this dissipative environment will have an important part in the appearence of a possible dispersion relation $\omega(k)$.

In connection with this problem we must add that the stability theory for solitons has many aspects depending upon the particular definition of the term ``stability'' and upon the mathematical methods been employed [18]. 

One approach in the investigation of stability is to assume a ``small perturbation'' from our solution and consider whether this 
perturbation grows with time. If the perturbation is small enough, the nonlinear equation that it obeys may be approximated by a 
linear equation. More powerful methods can then be used to decide whether the approximate linear equation implies growth or 
decay with time. 
Nevertheless, one must be very careful, because the drawn conclusions from a study of the linear equation may differ from the 
implications of the exact nonlinear equation that it approximates.

A second appoach is to employ the nonlinear stability theory, that was developed by Benjamin [19] in connection with the 
Korteweg-deVries equation. In this case the question of stability is answered, but the problem of ``asymptotic stability'' remains 
open. The issue of stability for our model will be addressed in a further paper. 

Concerning  QM there is another interesting problem and it addresses the \-cons\-truc\-tion\- of wave packets for classical objects 
out of quantum wave packets in such a way that the resultant wave packets are not correlated so that one can identify 
each resultant wave packet by a classical particle. There are two different approches in this direction [20].

In the first group are those models that assert that macroscopic bodies possess classical properties even if they are perfectly isolated, in other words, classical behaviour is intrinsic to macroscopic bodies. 
In this set we may find the model proposed by K\'arolyh\'azy [21], which affirms that there is a ``coherence cell'', the spatial domain inside which the superposition principle still holds for the center of mass wave function of an isolated solid body. 
When this ``coherence cell'' expands to a certain size, then a spontaneous, stochastic reduction on the wave function takes place. 
The origin of these reductions is ascribable to the uncertainty structure of empty-space, in other words, there are no external agents involved in these reductions, i.e. it is postulated that these reductions are a fundamental behaviour of quantum systems. 

On the other hand we may find some proposals which assert that classical properties are not intrinsic to 
macroscopic bodies. 
It is noteworthy to mention that in these works the main idea is, as Zeh has pointed out [22], 
macroscopic quantum systems are never isolated from their environments, and as a result quantum coherence 
``leaks out'' into the environment [23]. Even the possible relevance of a cosmic influence on the generation of this 
classical behaviour was considered by Wigner [24].
The role of gravity as a possible trigger of the needed  dynamical supression of linear superpositions of macroscopically 
distinguishable states has already been analyzed [25, 26] but, in order to avoid some inconsistencies that 
appear in some of these models [25], a new fundamental length [13] has to be introduced. 
The non-existence of a quantum theory of gravity could pose an additional difficulty to the study of gravity as a 
feasible candidate. 

In this direction,  it could be interesting to see if the Higgs field could play a similar role that in some of the previous attempts 
the gravitational field has played, and understand if this classical property could emerge by means of an interaction with the Higgs field. 

There is another question that remains unanswered, namely, the corresponding relativistic case of our situation. We will not address here 
this issue, but  some previous work [27] could be a good starting point for this quest. We might begin the analysis of this case using the relation between the transformation pro\-per\-ties\- of the fields and 
their corresponding motion equations [28] in order to derive two second order relativistic equations, each one of them 
depending only on $e_L$ or $e_R$, and afterwards try to look for a possible generalization of the results of Degasperis [8], which would then render localized solutions.   
\bigskip
\bigskip

\large{\bf Acknowlegdments.}\normalsize
\bigskip

This work has been partially supported by CONACYT Grant No. 3544--E9311. A. C. would like to thank Deutscher Akademischer Austauschdienst (DAAD) for the fellowship received. 
\vfil \eject  

\end{document}